\documentclass[fleqn,10pt]{wlpeerj}

\usepackage{float}

\usepackage{listings}
\title{Wave Sort (W-Sort): A Novel In-Place Algorithm with Dynamic Pivoting for Consistent Performance and Logarithmic Stack Depth}

\author[1]{Jia Xu Wei}
\affil[1]{University of California, Davis, USA}
\corrauthor[1]{Jia Xu Wei}{jxuwei@ucdavis.edu}

\begin{abstract}
Modern comparison sorts like quicksort suffer from performance inconsistencies due to suboptimal pivot selection, leading to \(O(N^2)\) worst-case complexity, while in-place merge sort variants face challenges with data movement overhead. We introduce Wave Sort, a novel in-place sorting algorithm that addresses these limitations through a dynamic pivot selection strategy. Wave Sort iteratively expands a sorted region and selects pivots from this growing sorted portion to partition adjacent unsorted data. This approach ensures robust pivot selection irrespective of dataset size, guarantees a logarithmic recursion stack depth, and enables efficient in-place sorting. Our analysis shows a best comparison complexity of \(N-1\), average comparison complexity close to \(\log_2(N)!\), and worst-case comparison complexity bounded by \(O(N(\log(N))^2)\) with a small constant factor, which could be reduced to \(O(N\log(N))\) with hybrid sorting. The algorithm can be easily expanded to be hybridized with other sorting algorithms. Experimental results demonstrate that Wave Sort requires significantly fewer comparisons than quicksort on average (approximately 24\% less) and performs close to the theoretical minimum \(\log_2(N)!\). Wave Sort offers a compelling alternative for applications demanding consistent, predictable, and in-place sorting performance.
\end{abstract}

\begin{document}

\flushbottom
\maketitle
\thispagestyle{empty}

\section*{Introduction}

The primary cause of inconsistent performance in modern quicksort implementations \citep{Sedgewick-quicksort} is lopsided partitions resulting from suboptimal pivot values. This can lead to a significant increase in both the number of comparisons and the depth of the recursion stack, degrading performance from an average of \(O(N\log N)\) to a worst-case of \(O(N^2)\). Due to the computational cost associated with selecting a perfectly balanced pivot, current implementations often employ heuristic methods, such as the "median of three" \citep{Sedgewick-quicksort}. Although these methods attempt to choose a pivot from a representative sample, their efficacy can diminish as the size of the data set increases because the sample size remains constant and does not scale accordingly. To ensure robust pivot selection, regardless of the size of the data set, the selection process would ideally need to be dynamic, considering an increasingly larger portion of the data. However, this approach typically increases computational overhead.

The standard merge sort \citep{Sedgewick-mergesort} guarantees a comparison time of \(O(N\log(N))\) and logarithmic stack depth, but it requires additional linear space \(O(N)\) and involves \(O(N\log(N))\) data movements. Although in-place merge sort variants \citep{in-place-mergesort} eliminate the need for additional space, the complexity of comparison and element movement often increases with higher constant factors \citep{inplace-merging}.

We introduce wave sort, a divide-and-conquer, in-place sorting algorithm designed to address the aforementioned shortcomings of quicksort and merge sort. The algorithm begins by sorting an initial small portion of the data. Then it utilizes the median value of this sorted portion as the pivot to partition and sort progressively larger segments of the data until the entire dataset is sorted. This strategy effectively expands the algorithm's "frame of reference" for selecting a balanced pivot by leveraging already sorted sections of the array. The algorithm is divided into two logic components, up-wave and down-wave, to simplify the expansion, such as integration with adaptive sorting or hybrid with other sorting algorithms. Each component could have different expansion approaches. This paper shows that adaptive sorting is integrated to handle the pre-sorted sequence. Other adaptive sorting approaches, such as those shown in MBISort\citep{MBISort}, could also be integrated. It is also easy to effectively trade off comparison for element movement. Although leap-frogging samplesort \citep{Leapfrogging-samplesort} and some variations such as partition sort \citep{partition-sort} employ a similar concept, their implementations lack efficient block swap mechanisms, which could lead to extra element moves. Furthermore, applying adaptive sorting techniques to these existing variations can be challenging as shown in PEsort\citep{PEsort}.

Experimental work indicates that, on average, wave sort uses significantly fewer comparisons than quicksort (approximately 24\% less) and a comparable number of element moves. The worst-case comparison complexity for wave sorting is \(O(N(\log(N))^2)\), notably with a small constant factor (approximately \(\frac{1}{4}\)). Adaptive sorting techniques are integrated to efficiently handle partially or fully sorted sequences. Additionally, the recursion stack depth is inherently restricted to logarithmic depth. Wave sort could also replace quick sort in Introsort \citep{introsort} to reduce the worst-case comparison complexity to \(O(N\log(N))\). Wave sort makes the judgment of an unbalanced partition more accurate, so that it can avoid falsely falling into slower heapsort too early. The wave sort has \(O(\log(N)^2\) stack depth in the worst case, so it is safe for Introsort to adapt more aggressive approaches. 

\section{Core Algorithm}

The algorithm is named \textbf{wave sort} because its active working set dynamically expands and contracts during execution, resembling the motion of a wave as it processes the data. The wave sort comprises two primary operations: \textbf{up-wave} and \textbf{down-wave}.
The \textbf{up-wave} operation is responsible for expanding the currently sorted region to incorporate an adjacent unsorted portion of the data. Once this expanded working set (containing both the established sorted part and the newly included unsorted part) is defined, it is passed to the \textbf{down-wave} operation for sorting.
The \textbf{down-wave} operation then sorts the unsorted portion within this working set using a divide-and-conquer approach. Crucially, pivots for partitioning the unsorted data are selected from the already sorted portion of the working set. The overall process is iterative:
\begin{enumerate}[noitemsep] 
    \item An up-wave phase identifies and incorporates new unsorted data adjacent to the existing sorted region.
    \item A subsequent down-wave phase sorts this newly included unsorted data, thereby enlarging the overall sorted region. This cycle repeats until the entire sequence is sorted.
\end{enumerate}
During a down-wave operation:
\begin{itemize}[noitemsep] 
    \item If the unsorted portion within the current working set becomes fully sorted, the down-wave operation for that set is complete. The process may then loop back to the up-wave operation to further expand the now larger sorted region.
    \item If the sorted portion (which acts as the pivot provider) is "exhausted" (all its elements have been used as pivots, or its length is too short for effective partitioning) before the current unsorted portion is fully sorted, the remaining unsorted data are handled by initiating a new up-wave operation. This new operation will establish a new sorted base to continue sorting the remaining.
\end{itemize}
To simplify the description and without loss of generality, the following assumptions are made:
\begin{enumerate}[noitemsep] 
    \item The array is to be sorted in \textbf{ascending order} (i.e., from the least value to the greatest, with the least value ultimately residing at the leftmost position of the sorted array).
    \item The sorting process starts at the right end. An initial small segment at this end is sorted first, and then the "wave" of sorting progresses towards the other end.
\end{enumerate}
For brevity, we will be using the Go programming language \cite{programming:lang:go} to illustrate the wave sort. The code complies with version 1.18 and above.

\subsection{Driver function}
This function serves as the initial entry point, handling the base case for sequences with fewer than two elements by simply returning. For sequences of two or more, it delegates the core sorting logic to the core algorithm.
\begin{lstlisting}[caption=Driver function, label=lst:driver]
type seq []int

func (s seq) WaveSort() {
    if len(s) < 2 {
	return
    }
    s.upwave(0, len(s)-1)
}
\end{lstlisting}

\subsection{up-wave}

The wave sort commences by establishing an initial sorted segment within the array. Analogously to dynamic programming, this establishes a base case or solves an initial "subproblem." This initial segment can be as small as a single element (which is inherently considered sorted) or a larger portion sorted using a conventional algorithm.

Following the establishment of this initial sorted segment, the \textbf{up-wave} operation takes charge. Its primary role is to iteratively expand the bounds of the active working set to incorporate adjacent unsorted data. After each expansion, the up-wave operation passes the newly defined working set (comprising the current sorted region and the newly included unsorted data) to the \textbf{down-wave} operation for sorting.

\begin{figure}[H]
    \centering
    \includegraphics[width=1\linewidth]{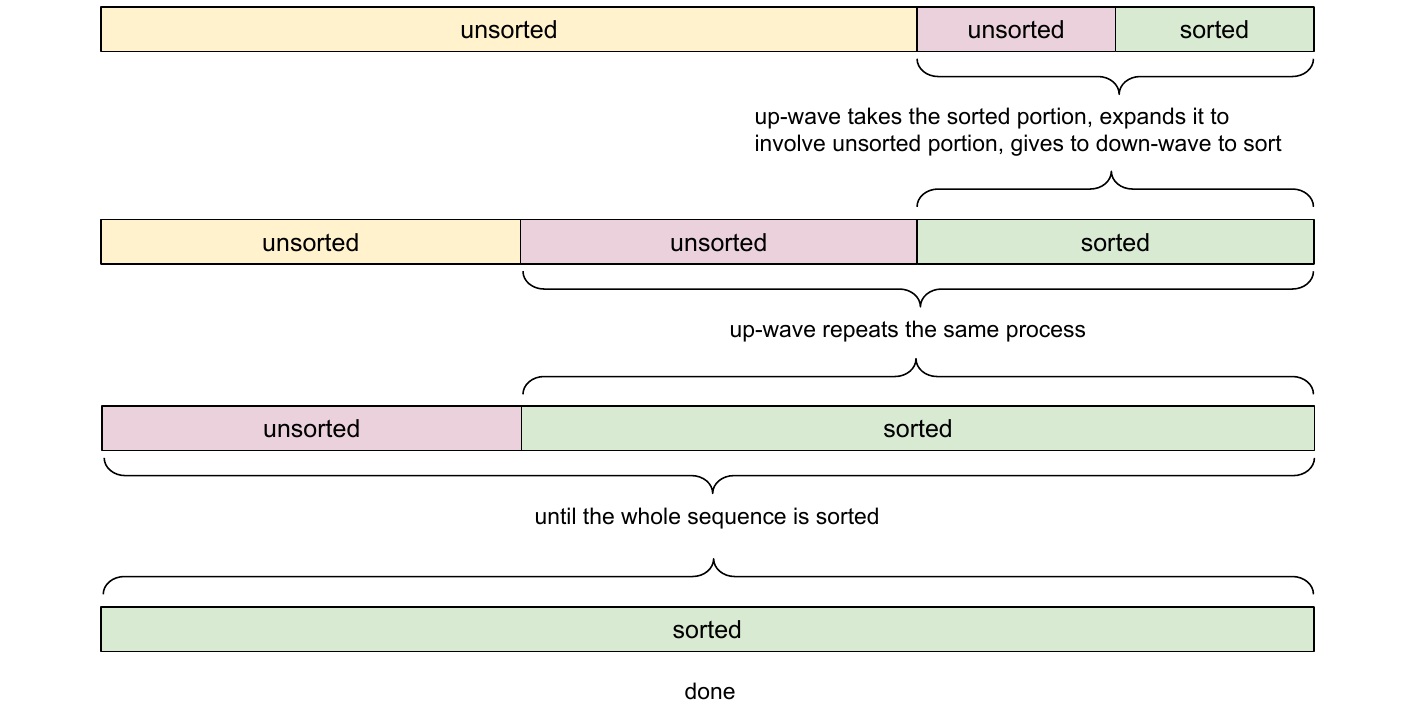}
    \caption{up-wave}
    \label{fig:up-wave}
\end{figure}

Once the down-wave completes sorting this set, thereby enlarging the overall sorted region, the up-wave operation re-engages to expand the working set further. This cycle continues until the entire array is sorted. The up-wave thus serves as the main driver and entry point for the overall wave-sorting process. The mechanics of the up-wave process are illustrated in Fig.~\ref{fig:up-wave}, and Listing \ref{lst:upwave}.

\begin{lstlisting}[caption=Up-wave Algorithm, label=lst:upwave]
func (s seq) upwave(start, end int) {
    if start == end {
	return
    }
    sortedStart := end
    sortedLength := 1
    leftBound := end - 1
    length := end - start + 1
    for {
	s.downwave(leftBound, sortedStart, end)
	sortedStart = leftBound
	sortedLength = end - sortedStart + 1
	if length < sortedLength<<2 {
		break
	}
	leftBound = end - sortedLength<<1 + 1
    }
    s.downwave(start, sortedStart, end)
}
\end{lstlisting}

\subsection{down-wave}

Before diving into the details of the down-wave, it is necessary to introduce two subprocesses: partitioning and block swapping.

\subsubsection{Partitioning}

Partitioning splits the sequence into two parts: The left part contains elements less than the pivot, and the right part contains elements greater than the pivot. The function returns the index of the split point that lies within the greater part.

\begin{figure}[H]
    \centering
    \includegraphics[width=1\linewidth]{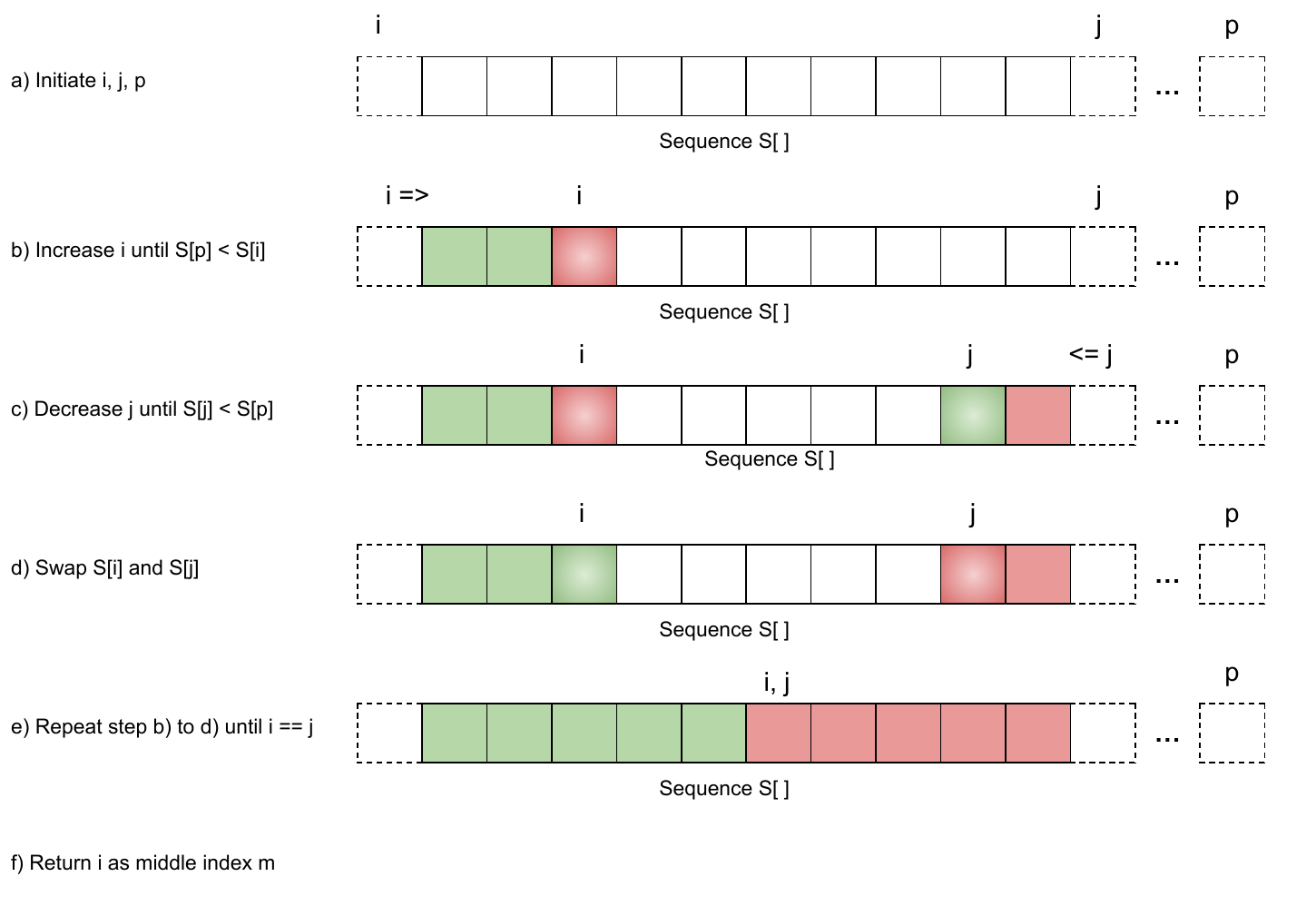}
    \caption{Partitioning}
    \label{fig:partitioning}
\end{figure}

Initialize the pointer \verb|i| to the index of the position before the first element, and the pointer \verb|j| to the index of the position after the last element. Let \verb|p| be the pivot. Increment \verb|i| until the element at index \verb|i| is greater than or equal to the pivot. If \verb|i == j|, return \verb|i| as the middle index \textit{m}. Then decrement \verb|j| until the element at index \verb|j| is less than the pivot. If \verb|j == i|, return \verb|i| as the middle index \textit{m}. Then swap the elements at the indices \verb|i| and \verb|j|. Repeat this process. The partitioning process is illustrated in Fig.~\ref{fig:partitioning} and Listing ~\ref{lst:partition}. 

The partitioning algorithm is different from the one from \cite{Sedgewick-quicksort} and many other standard libraries that takes n+1 comparisons on an array of size n. It only takes n-1 comparisons on the same array and can be used to replace these algorithms to improve performance. The quicksort used to compare with the wave sort in this paper uses the same partitioning algorithm as the wave sort to eliminate unnecessary comparisons.

\begin{lstlisting}[caption=Partitioning Algorithm, label=lst:partition]
func (s seq) partition(l, r, p int) int {
    i, j := l-1, r
    for {
	for {
		i++
		if i == j {
			return i
		}
		if s.less(p, i) {
			break
		}
	}
	for {
		j--
		if j == i {
			return i
		}
		if s.less(j, p) {
			break
		}
	}
	s.swap(i, j)
    }
}
\end{lstlisting}

\subsubsection{Block swap}

Partitioning divides the unsorted portion into left and right blocks, separated by index \textit{m}. The sorted portion is also divided into left and right blocks, separated by the pivot element. In the block-swapping process, the right block of the unsorted portion (which becomes the left block in the swap) is exchanged with the left block of the sorted portion (which becomes the right block in the swap).

If the left block is shorter than the right block, define the length of the left block as \(ll= r-m\), where \textit{r} is the index of the first element in the right block and \textit{m} is the index of the first element in the left block. Set \(nm = p-ll+1\), and initialize \(j=m\). Store the element at index \textit{m} in a temporary variable. Move the element of index \(j+ll\) to index \textit{j}, then update \textit{j} to \(j+ll\). If \(j \geq nm\), replace the element on the index \(j+ll\) with the one on the index \(j-nm+m\). Repeat this process for \(p-m+1\) iterations, as shown in Listing ~\ref{lst:blockSwap_sl}.
\begin{lstlisting}[caption=Short left, label=lst:blockSwap_sl]
func (s seq) blockSwap_sl(m, p, ll int) {
    tmp := s[m]
    init := m
    j := m
    nm := p - ll + 1
    var k int
    for range p - m + 1 {
	if j >= nm {
		k = j - nm + m
		if k == init {
			init++
			s[j] = tmp
			j = init
			tmp = s[j]
			continue
		}
	} else {
		k = j + ll
	}
	s[j] = s[k]
	j = k
    }
}
\end{lstlisting}

In this case, each element of the left and right blocks moves only once. The usage of the temporary variable is minimized. The total movement of the element is about the length of the left block \textit{ll} plus the length of the right block \textit{lr}. In a normal swap operation, there are 3 element moves. So each swap equals about 3 block swaps. In the case where the temporary variable could be squeezed into the register and the compiler could optimize the code to do so, there are two element moves in the swap. So each swap equals 2 block swaps in this case. In \cite{Leapfrogging-samplesort} and its variation \cite{partition-sort} to swap the left and right blocks, \textit{lr} times swaps are required, which equals \(2 \times lr\) element moves or \(3 \times lr\) element moves if the register variable is not applicable. The block swap in this paper only needs \(lr + ll\) element moves. For \(ll < lr\), the moves of the element are reduced even if the register variable is applicable.

If the left block is not shorter than the right one, swap elements between the right and left blocks one by one. Only a portion of the unsorted block, equal in size to the sorted block, is moved, as shown in Listing ~\ref{lst:blockSwap}.

\begin{lstlisting}[caption=Short right, label=lst:blockSwap_sr]
func (s seq) blockSwap_sr(m, r, p int) {
    i := m
    tmp := s[i]
    j := r
    for j < p {
	s[i] = s[j]
	i++
	s[j] = s[i]
	j++
    }
    s[i] = s[j]
    s[j] = tmp
}
\end{lstlisting}

In this case, the temporary variable is used only once. The moves of the element are the same as \cite{Leapfrogging-samplesort} and its variation \cite{partition-sort}  if the register variable is applicable. Otherwise, the moves are reduced to \(\frac{2}{3}\).

Together, the two scenarios are decided by the function shown in Listing ~\ref{lst:blockSwap}.

\begin{lstlisting}[caption=blockSwap, label=lst:blockSwap]
func (s seq) blockSwap(m, r, p int) {
    ll := r - m
    if ll == 0 {
	return
    }
    lr := p - r + 1
    if lr <= ll {
	s.blockSwap_sr(m, r, p)
	return
    }
    s.blockSwap_sl(m, p, ll)
}
\end{lstlisting}

\begin{figure}[H]
    \centering
    \includegraphics[width=1\linewidth]{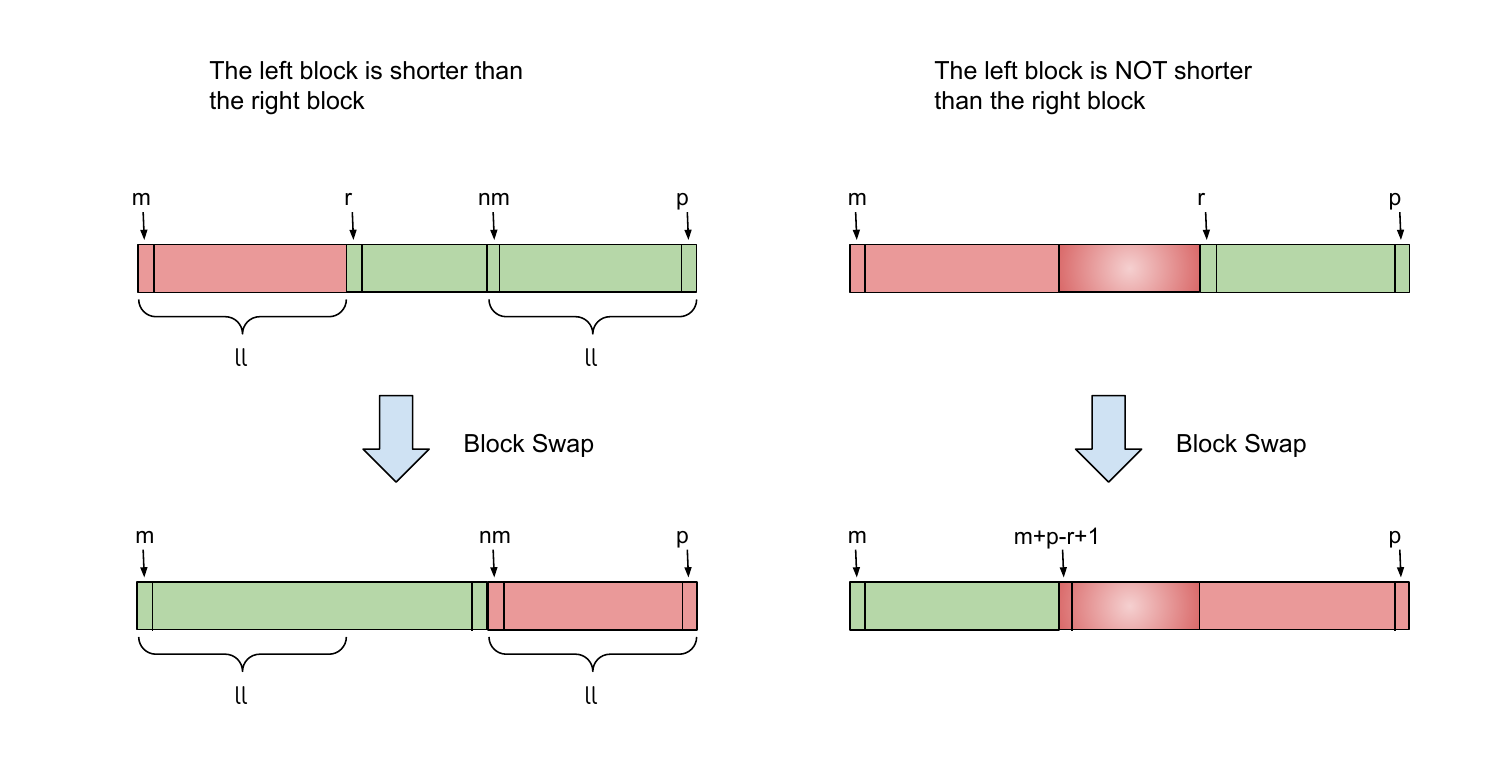}
    \caption{block swapping}
    \label{fig:block-swapping}
\end{figure}

In both cases, each element concerned is moved only once, with minimal use of a temporary variable. Only the elements concerned are moved. This ensures maximum efficiency and eliminates the noise introduced in \cite{Leapfrogging-samplesort} and its variation \cite{partition-sort} for further analysis. The block swap process is illustrated in Fig.~\ref{fig:block-swapping}.

\subsubsection{Down-wave process}\hfill\\

The down-wave selects the median of the sorted portion as the pivot and performs partitioning on the unsorted portion. This partition returns a middle index \( m \), which acts as the delimiter between two subsets. Since the pivot corresponds to the median of the sorted portion, the partition ensures that the elements in the left and right subsets of the unsorted section fall on the correct sides of the pivot in the final sorted array.

Specifically, the elements to the left of \( m \) in the unsorted portion, along with those to the left of the pivot in the sorted portion, belong to the left of the pivot in the final output. In contrast, the elements to the right of \( m \) in the unsorted portion and to the right of the pivot in the sorted portion belong to the right of the pivot. A block swap is performed to exchange the right subset of the unsorted portion with the left subset of the sorted portion. This results in two new sections, each composed of an unsorted part on the left and a sorted part on the right.

\begin{figure}[H]
    \centering
    \includegraphics[width=1\linewidth]{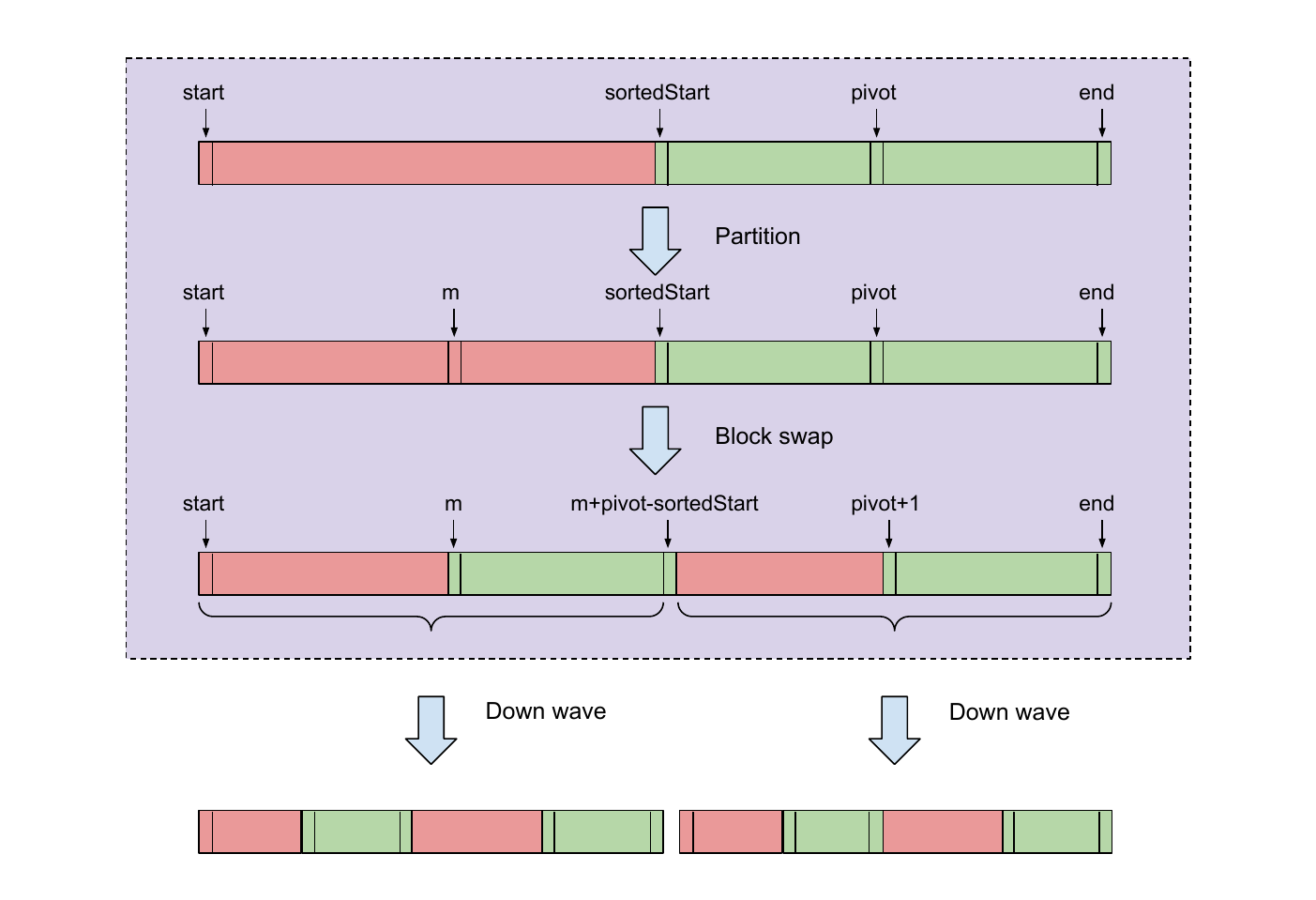}
    \caption{down wave}
    \label{fig:down-wave}
\end{figure}

The same partition and block swap operations are applied recursively to these new sections, progressively forming smaller subsections with the same structure. This continues until either the sorted or unsorted portion is fully exhausted. If the sorted portion is exhausted, an up-wave can be initiated to process the remaining unsorted portion, starting from its rightmost elements. If the unsorted portion is exhausted, the sorting process is complete. The down-wave procedure is illustrated in Fig.~\ref{fig:down-wave} and in Listing ~\ref{lst:downWave}.

\begin{lstlisting}[caption=Down-wave, label=lst:downWave]
func (s seq) downwave(start, sortedStart, end int) {
    if sortedStart-start == 0 {
	return
    }
    p := sortedStart + (end-sortedStart)/2
    m := s.partition(start, sortedStart, p)
    if m == sortedStart {
	if p == sortedStart {
		s.upwave(start, sortedStart-1)
		return
	}
	s.downwave(start, sortedStart, p-1)
	return
    }
    s.blockSwap(m, sortedStart, p)
    if m == start {
	if p == sortedStart {
		s.upwave(m+1, end)
		return
	}
	p++
	s.downwave(m+p-sortedStart, p, end)
	return
    }
    if p == sortedStart {
	s.upwave(start, m-1)
	s.upwave(m+1, end)
	return
    }
    s.downwave(start, m, m+p-sortedStart-1)
    s.downwave(m+p-sortedStart+1, p+1, end)
}
\end{lstlisting}

\section{Analysis}

Let the length of the sequence be \( n \), and let the length of the sorted portion be \( \mathit{ls} \). If \( \mathit{ls} \) is odd, it can be expressed as \(2m+1\). The probability that the median of the sorted portion appears at position \( i \) in the sequence is:
\begin{equation}
p(i) = \frac{\binom{i - 1}{m} \cdot \binom{n - i}{m}}{\binom{n}{2m + 1}}
\end{equation}
If \( \mathit{ls} \) is even, it can be expressed as \( 2m \). In this case, the probability that the median appears at position \( i \) is:
\begin{equation}
p(i) = \frac{\binom{i - 1}{m-1} \cdot \binom{n - i}{m}}{\binom{n}{2m}}
\end{equation}
The probability of evenly dividing the unsorted portion is greater than \( \frac{1}{n} \), as is the case in quick sort. Therefore, the average number of comparisons is expected to be closer to the best case than in quicksort.
To simplify the analysis, assume the expansion ratio of the up-wave is always 2; that is, each up-wave step doubles the size of the working set from the sorted portion. In the best case, the total number of comparisons for the largest up-wave working set is bounded by \( C_0 \):
\[
C_0 = \frac{n}{2} + \left(\frac{n}{4} + \frac{n}{4}\right) + \cdots  - E_0= \frac{n}{2} \log_2 \frac{n}{2} - E_0
\]
The difference \(E_0\) is that in each division step of the down wave, the pivot is removed from the sorted subset. So some of the branches may end earlier and the comparison only needs to be done among about half of the pairs that are approximately equal to \(\frac{n}{8}\) or a little more.
The comparisons for the next smaller working set are bounded by \( C_1 \):
\[
C_1 = \frac{n}{4} \log_2 \frac{n}{4} - E_1
\]
where \(E_1 = E_0/2 \approx \frac{n}{16}\).
Thus, the total comparison is \(C_b\): 
\[C_b = C_0+C_1+ ...C_m - E = nlog_2n-(n-1)-E\]
where \(m = log_2\frac{n}{2}-1\) , \(E = \sum_{i=0}^{m-1}E_i \approx \frac{n}{4}\). 
\begin{align*}
C_b \approx nlog_2n - (n-1) - \frac{n}{4} \\ = nlog_2n - \frac{5n}{4} + 1
\end{align*}
In the worst case, each comparison provides the minimum information (lowest entropy). A sorted sequence fulfills this situation, so a sorted sequence is used for analysis and testing. The total number of comparisons is denoted by \(C_w\).
In this case, the number of comparisons for the largest up-wave working set is \(\frac{n}{2}\log_2\frac{n}{2}\) and there is only one such working set. The comparisons for the next smaller working set are \(\frac{n}{4}\log_2\frac{n}{4}\) and the number of such working sets is 2. So we get:
\begin{align*}
C_w = 1\times\frac{n}{2}\log_2\frac{n}{2} + 2\times\frac{n}{4}\log_2\frac{n}{4} + 4\times\frac{n}{8}\log_2\frac{n}{8} \\ + \cdots + \frac{n}{8}\times4\log_24 + \frac{n}{4}\times2\log_22 \\
= \frac{n}{4} \left( \log_2^2 n - \log_2 \frac{n}{4} \right)
\end{align*}
The maximum stack depth is approximately \( \log_2 n \) as in the quick sort. In the worst case, the stack depth is less than:
\[
\frac{1}{2} \left( \log_2^2 n + \log_2 n \right)
\]

\section{Experimental Results}

The experimental code was run on randomly shuffled sequences and compared to quick sort. The sequence is shuffled with the \verb|Shuffle| function from the go package "math/rand/v2". The quick sort is the improved version with the partition function proposed in this paper which reduces comparisons and keeps the same swaps. The results demonstrate that the basic wave sort algorithm performs close to the theoretical minimum comparison limit of $\log_2(n!)$. Across multiple experiments, wave sorting required approximately \(1\%\) more comparisons than this theoretical limit on average. This efficiency gap decreases as the sequence length increases (see Table \ref{tab:average_comparison}). The code is shown in the Appendix \ref{appendix:wavesort}.

\begin{table}[H]
    \centering
    \begin{tabular}{|c|c|c|c|l|} \hline 
         sequence size, $n$&  wave sort&  quick sort& $\log(n!)$ &$C_b$\\ \hline 
         $2^{20}$&  19,680,967&  26,070,556& 19,458,756 &19,660,801\\ \hline 
         $2^{21}$&  41,459,069&  55,091,250& 41,014,653 &41,418,753\\ \hline 
         $2^{22}$&  87,113,082&  115,662,060& 86,223,598 &87,031,809\\ \hline 
         $2^{23}$&  182,613,766&  243,388,480& 180,835,793 &182,452,225\\ \hline
    \end{tabular}
    \caption{Average Comparison}
    \label{tab:average_comparison}
\end{table}

Although wave sort requires fewer comparisons than quick sort, it performs more data swaps and requires additional block swaps. In wave sort, each block swap moves each element once and minimizes access to the temporary variable to only once in most cases. Each swap is equivalent to approximately 3 block swaps. The total data movement in the basic wave sort is more than twice that in the quick sort (see Table \ref{tab:average_swap}). This issue is addressed in trade-off comparisons with swaps.

\begin{table}[H]
    \centering
    \begin{tabular}{|c|c|c|c|} \hline 
         sequence size, $n$&  wave sort, swap&  wave sort, block swap& quick sort, swap\\ \hline 
         $2^{20}$&  5,001,848&  16,984,185& 4,757,387\\ \hline 
         $2^{21}$&  10,528,024&  36,064,105& 9,994,835\\ \hline 
         $2^{22}$&  22,104,628&  76,317,658& 20,976,731\\ \hline 
         $2^{23}$&  46,307,301&  161,022,256& 43,869,539\\ \hline
    \end{tabular}
    \caption{Average Swap}
    \label{tab:average_swap}
\end{table}

In practice, many sequences encountered are presorted or partially presorted. Wave sort exhibits its worst-case performance when processing fully presorted sequences (see Table \ref{tab:worst_comparison}). This issue is addressed in adaptive and hybrid sorting.

\begin{table}[H]
    \centering
    \begin{tabular}{|c|c|c|} \hline 
         sequence size, $n$&  wave sort& $C_w$\\ \hline 
         $2^{20}$&  100,139,008& 100,139,008\\ \hline 
         $2^{21}$&  221,249,536& 221,249,536\\ \hline 
         $2^{22}$&  486,539,264& 486,539,264\\ \hline 
         $2^{23}$&  1,065,353,216& 1,065,353,216\\ \hline
    \end{tabular}
    \caption{Worst Comparison}
    \label{tab:worst_comparison}
\end{table}

\section{Trade off Comparisons with Swaps}

The average comparison count of wave sort is quite close to the theoretical limit, leaving only a small margin for improvement. In practice, comparisons usually cost more than swaps on complex data structures, as comparisons require examining the content of the data structure, while swaps only require exchanging pointers to the content. Therefore, in general, fewer comparisons are preferred. However, in some cases, trading comparisons for swaps may be advantageous.

Since wave sort performs more than twice the swaps compared to quick sort, there is room for optimization if such trade-offs prove beneficial. The basic strategy is to increase the expansion rate in the up-wave so that partitioning works more effectively to reduce swaps. In both down and up-wave processes, sorting short sequences with insertion sort can replace swaps with block swaps for greater efficiency. The smaller sorted portions also reduce the number of block swaps required. An implementation showing these trade-offs is present in Listing ~\ref{lst:trade-off}. 

\begin{lstlisting}[caption=Trade-off Comparison, label=lst:trade-off]
func (s seq) upwave(start, end int) {
	length := end - start + 1
	if length == 1 {
		return
	}
	if length <= 12 {
		s.insertSort(start, end)
		return
	}
	sortedStart := end - 7
	s.insertSort(sortedStart, end)
	sortedLength := 8
	leftBound := end - 127
	for leftBound > start {
		s.downwave(leftBound, sortedStart, end)
		sortedStart = leftBound
		sortedLength = end - sortedStart + 1
		if length <= sortedLength<<6 {
			break
		}
		sortedLength <<= 4
		leftBound = end - sortedLength + 1
	}
	s.downwave(start, sortedStart, end)
}

func (s seq) downwave(start, sortedStart, end int) {
	unsortedLength := sortedStart - start
	if unsortedLength == 0 {
		return
	}
	if end - sortedStart == 0 {
		if unsortedLength < 12 {
			s.insertSort(start, end)
			return
		}
		s.upwave(start, end)
		return
	}
	p := sortedStart + (end-sortedStart)/2
    ...
}

func (s seq) insertSort(l, r int) {
	sortedIndex := r - 1
	for ; sortedIndex >= l; sortedIndex-- {
		low := sortedIndex
		hi := r
		m := (low + hi + 1) >> 1
		for low < hi {
			if s.less(sortedIndex, m) {
				hi = m - 1
				m = (low + hi + 1) >> 1
				continue
			}
			low = m
			m = (low + hi + 1) >> 1
		}
		if m > sortedIndex {
			tmp := s[sortedIndex]
			for j := sortedIndex; j < m; j++ {
				s[j] = s[j+1]
			}
			s[m] = tmp
			bs += m - sortedIndex + 1
		}
	}
}
\end{lstlisting}

A typical trade-off result is shown in Table~\ref{tab:trade_off} with the implementation demonstrated in the above code. The comparison increases approximately \(1.6\%\). One comparison trades for approximately 2.6 swaps plus 46 block swaps, which is equivalent to approximately 25 swaps (see Table \ref{tab:trade_off}). The code is shown in the Appendix \ref{appendix:trade-off}. 
More detailed mathematical analysis is required to determine the optimal expansion rate. Generally, the longer the sorted portion, the larger the expansion rate should be, based on the probability equation provided above.
\begin{table}[H]
    \centering
    \begin{tabular}{|c|c|c|c|} \hline 
         sequence size, $n$&  comparison&  swap& block swap\\ \hline 
         $2^{20}$&  20,105,673&  4,235,207& 4,252,992\\ \hline 
         $2^{21}$&  42,359,965&  8,974,366& 9,777,278\\ \hline 
         $2^{22}$&  87,822,911&  19,070,908& 22,700,097\\ \hline 
        $ 2^{23}$&  185,053,654&  40,244,980& 36,335,453\\ \hline
    \end{tabular}
    \caption{Trade off Comparison for Swap}
    \label{tab:trade_off}
\end{table}

\section{Adaptive and Hybrid Sorting}
For sorted or partially sorted sequences, regardless of whether they are reversed, an additional check is incorporated into the up-wave to identify the longest presorted sequence as the sorted portion. For a fully sorted sequence, this results in only \(n-1\) comparisons. If the sequence is completely reversed, the number of swaps is reduced to \(\frac{n}{2}\). This modification effectively short-circuits the processing of presorted sequences (see Listing ~\ref{lst:AdaptiveWave} and Table \ref{tab:adaptive_presorted}). The code is shown in the Appendix \ref{appendix:adaptiv}.

\begin{lstlisting}[caption=Adaptive Wave Sort, label=lst:AdaptiveWave]
func (s seq) upwave(start, end int) {
	length := end - start + 1
	if length == 1 {
		return
	}
	if length <= 12 {
		s.insertSort(start, end, end)
		return
	}
	sortedStart := s.preSorted(start, end)
	sortedLength := end - sortedStart + 1
	if sortedLength < 8 {
		s.insertSort(end-7, sortedStart, end)
		sortedStart = end - 7
		sortedLength = 8
	}
	sortedLength <<= 4
	leftBound := end - sortedLength + 1

	for leftBound > start {
		s.downwave(leftBound, sortedStart, end)
		sortedStart = leftBound
		sortedLength = end - sortedStart + 1
		if length <= sortedLength<<6 {
			break
		}
		sortedLength <<= 4
		leftBound = end - sortedLength + 1
	}
	s.downwave(start, sortedStart, end)
}

func (s seq) preSorted(start, end int) int {
	du := false

	for i := end; i > start; i-- {
		if du {
			if s.less(i, i-1) {
				return i
			}
			continue
		}
		if s.less(i-1, i) {
			if i == end {
				du = true
				continue
			}
			s.reverse(i, end)
			return i
		}
	}
	if du {
		return start
	}
	s.reverse(start, end)
	return start
}

func (s seq) reverse(start, end int) {
	i := start
	j := end
	for range (end - start + 1) / 2 {
		s.swap(i, j)
		i++
		j--
	}
}

func (s seq) insertSort(l, sortedStart, r int) {
	sortedIndex := sortedStart - 1
	for ; sortedIndex >= l; sortedIndex-- {
		i := sortedIndex
		for ; i < r; i++ {
			if s.less(sortedIndex, i+1) {
				break
			}
		}
		if i > sortedIndex {
			tmp := s[sortedIndex]
			for j := sortedIndex; j < i; j++ {
				s[j] = s[j+1]
			}
			s[i] = tmp
		}
	}
}
\end{lstlisting}

\begin{table}[H]
    \centering
    \begin{tabular}{|c|c|c|} \hline 
         sequence size, $n$&  wave sort adaptive& reverse swap\\ \hline 
         $2^{20}$&  1,048,575& 524,288\\ \hline 
         $2^{21}$&  2,097,151& 1,048,576\\ \hline 
         $2^{22}$&  4,194,303& 2,097,152\\ \hline 
         $2^{23}$&  8,388,607& 4,194,304\\ \hline
    \end{tabular}
    \caption{Adaptive Presorted}
    \label{tab:adaptive_presorted}
\end{table}

As the adaptive approach adds one more step to check the boundary of the sorted sequence, the number of comparisons increases slightly. This approach trades off swaps for block swaps. The comparisons increase by approximately \(0.15\%\). The number of swaps increases by approximately \(0.53\%\), while the block swaps decrease by approximately \(1.6\%\) (see Table \ref{tab:adaptive_average}). 

\begin{table}[H]
    \centering
    \begin{tabular}{|c|c|c|c|} \hline 
         sequence size, $n$&  comparison&  swap& block swap\\ \hline 
         $2^{20}$&  20,140,905&  4,261,710& 4,170,009\\ \hline 
         $2^{21}$&  42,416,613&  9,017,588& 9,642,118\\ \hline 
         $2^{22}$&  87,949,166&  19,166,119& 22,400,068\\ \hline 
         $2^{23}$&  185,336,094&  40,456,208& 35,665,233\\ \hline
    \end{tabular}
    \caption{Adaptive Average}
    \label{tab:adaptive_average}
\end{table}

The up-wave could adopt other adaptive sorting to short-circuit certain sequence patterns, such as shown in MBISort \cite{MBISort}. The best adaptive sorting needs further study to ensure that it does not introduce too many extra comparisons or swaps in not-covered patterns. 

\begin{figure}[H]
    \centering
    \includegraphics[width=1\linewidth]{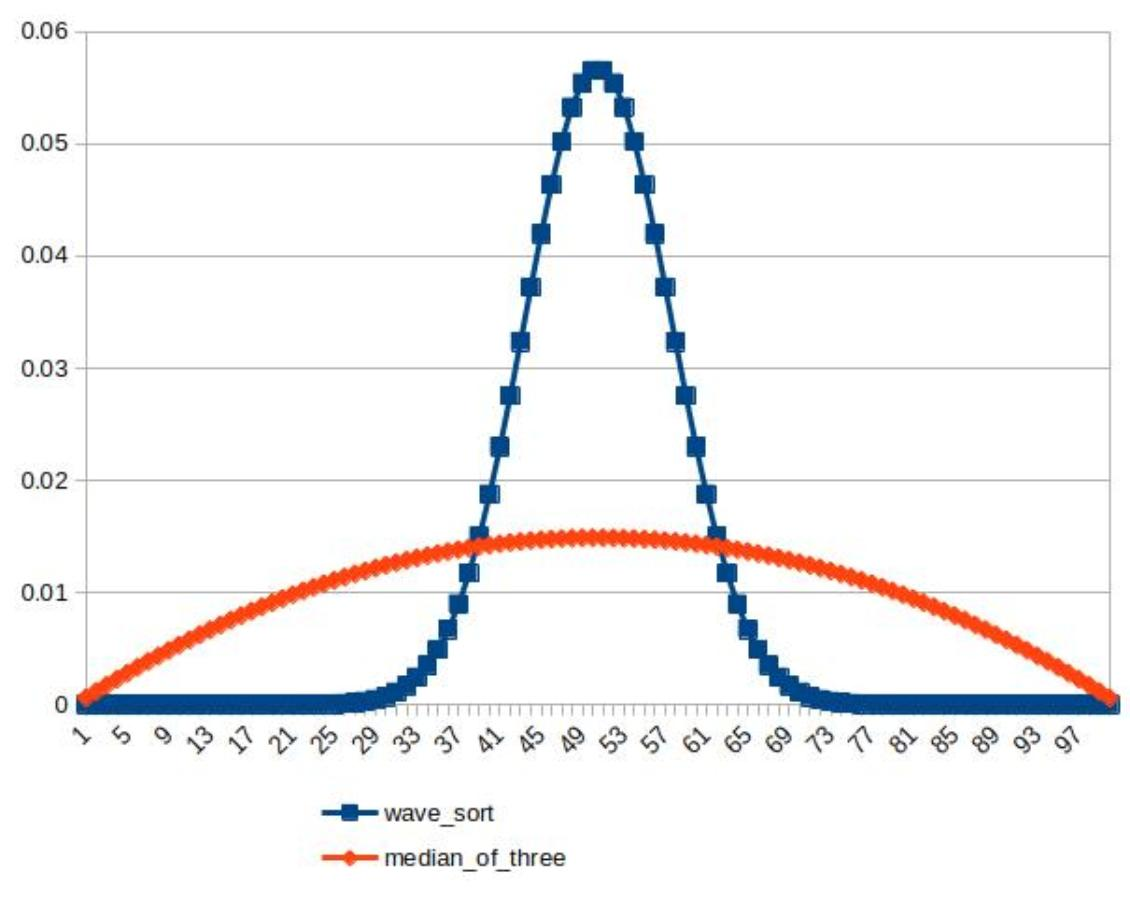}
    \caption{pivot-distribution}
    \label{fig:pivot-distribution}
\end{figure}

As adaptive sorting cannot cover all the possible sequence patterns, there are still sequence patterns for the wave sort to hit the worst case. The heap sort could be taken as the last resort, as in Introsort \cite{introsort} and pdqsort \cite{pdqsort} during the down-wave operation to ensure \(O(N\log(N))\) comparisons in the worst case. From the distribution of pivot position, as shown in figure \ref{fig:pivot-distribution} (it demonstrates the probability of pivot position in the case of 99 elements plus the pivot element. The wave sort has 99 samples and the median of three has 3 samples in this case), the probability of pivot position in wave sorting in the range of 1 to \(\frac{1}{8}n\) is very low, while the probability of median of three in the same range as in pdqsort \cite{pdqsort} is much higher. So in wave sort, it is quite unlikely that it falsely falls in the heap sort too early. It is also safe for Introsort to take more aggressive approaches to switch to heap sort earlier if there are unbalanced partitions occurring.

\section{Conclusion}

In this paper, we introduced Wave Sort, a novel divide-and-conquer in-place sorting algorithm designed to address key limitations found in widely used comparison sorts. Using a dynamic pivot selection strategy that leverages an expanding sorted region within the array, Wave Sort overcomes the pivot inconsistency issues that can lead to $O(N^2)$ worst-case performance in quicksort. Unlike standard merge sort, Wave Sort achieves sorting in place, avoiding the need for \(O(N)\) auxiliary space, and ensures a consistent, low logarithmic recursion stack depth. We refined the partition algorithm used by many standard quicksort libraries to reduce comparisons from \(N+1\) to \(N-1\), the theoretical limit, on an array of size \(N\). The performance of quick sorting is improved. We proposed a novel block-swap algorithm to reduce the movement of elements to the theoretical limit. It could be modified to replace the \textit{rotating swap} mentioned in \cite{in-place-mergesort}. The block swap performs \(\frac{a+b}{2}\) swaps on arrays of sizes \(a\) and \(b\). The rotating swap may perform \(a+b\) swaps on the same arrays. With all the extra operations removed, it helps to improve current standard libraries and performance analysis.

The algorithm is divided into two logic components, the up- and down-wave, each of which could have different expansion approaches, such as integration with adaptive sorting or hybrid with other sorting algorithms. 

Our analysis indicates a worst-case comparison complexity of \(O(N(\log(N))^2)\) with a small constant factor ($\approx \frac{1}{4}$). Experimental evaluation of randomly shuffled data demonstrated that Wave Sort achieves an average comparison count significantly lower than quicksort (approximately \(24\%\) less) and remarkably close to the theoretical minimum of \(\log(N)!)\), aligning with our analytical expectations. Furthermore, the algorithm is readily adaptable to handle partially or fully presorted and reversed sequences with high efficiency, achieving linear-time complexity in these best-case scenarios through integrated adaptive improvement. While the basic implementation involves more data movement operations than quicksort, we showed that the algorithm's structure allows for effective trade-offs to reduce swaps by increasing comparison operations, depending on the relative costs in a given application context. Running on a laptop with the comparison of integers, which is the fastest data type to compare in most CPU architectures and is implemented in the Go programming language, sorting a one million integer sequence takes about \(12.96\) seconds for 100 runs. The improved quicksort takes about \(11.71\) seconds. After increasing the complexity of the comparison, the wave sort takes about \(120\) seconds and the improved quicksort takes about \(140\) seconds. As the code is mainly for showcasing the algorithm and is not optimized for performance, and since Go is not an ideal language for heavy lifting jobs, the running time is just for reference.

Wave Sort presents a compelling alternative for scenarios that require efficient in-place sorting with predictable performance characteristics. Its combination of guaranteed logarithmic stack depth, favorable average comparison performance, and bounded worst-case behavior positions it as a strong candidate for replacing algorithms like in-place merge sort variants and the quick sort in Introsort in standard library implementations, especially where robust and consistent sorting is paramount.

Future work includes conducting a more detailed formal analysis to determine the optimal parameters for trading comparisons against element moves, thereby maximizing practical performance. Further research could also explore variations in adaptive sorting, hybrid sorting in Introsort, and its applicability to related sorting problems or data structures. Wave Sort represents a promising direction for advancing the state-of-the-art in in-place comparison sorting by effectively balancing theoretical elegance with practical performance considerations.

\section*{Acknowledgments}

To my father, Wei Wei, for inspiring me to write this in the first place.

\bibliography{sample}

\appendix
\section{wave sort - basic}\label{appendix:wavesort}
\begin{lstlisting}[caption=Wave Sort Basic Implementation, label=lst:wavesort_basic]

package wave_sort

type seq []int

var compare, swap, blockSwap int

// partition splits the sequence into two parts, the left part is less than the pivot and the right part is greater than the pivot.
// It returns the index of the split point which is involved in the greater part.
// l is the index of the left bound of the unsorted part.
// r is the index of the right bound of the unsorted part which is the firt element of the sorted part.
// p is the index of the pivot.
func (s seq) partition(l, r, p int) int {
	i, j := l-1, r

	for {
		for {
			i++
			if i == j {
				return i
			}
			if s.less(p, i) {
				break
			}
		}
		for {
			j--
			if j == i {
				return i
			}
			if s.less(j, p) {
				break
			}
		}
		s.swap(i, j)
	}
}

func (s seq) less(i, j int) bool {
	compare++
	return s[i] < s[j]
}

// blockSwap swaps the block from m to r and the one from r to p.
// m is the index of the first element of the left block.
// r is the index of the right bound of the left block and the first element of the right block.
// p is the index of the last element of the right block.
func (s seq) blockSwap(m, r, p int) {
	ll := r - m
	if ll == 0 {
		return
	}

	lr := p - r + 1

	if lr == 1 {
		s[m], s[p] = s[p], s[m]
		swap++
		return
	}

	if lr <= ll {
		s.blockSwap_sr(m, r, p)
		blockSwap += lr << 1
		return
	}

	s.blockSwap_sl(m, p, ll)
	blockSwap += ll + lr
}

func (s seq) blockSwap_sl(m, p, ll int) {
	tmp := s[m]
	init := m
	j := m
	nm := p - ll + 1
	var k int
	for range p - m + 1 {
		if j >= nm {
			k = j - nm + m
			if k == init {
				init++
				s[j] = tmp
				j = init
				tmp = s[j]
				continue
			}
		} else {
			k = j + ll
		}
		s[j] = s[k]
		j = k
	}
}

func (s seq) blockSwap_sr(m, r, p int) {
	i := m
	tmp := s[i]
	j := r
	for j < p {
		s[i] = s[j]
		i++
		s[j] = s[i]
		j++
	}
	s[i] = s[j]
	s[j] = tmp
}

func (s seq) swap(i, j int) {
	swap++
	s[i], s[j] = s[j], s[i]
}

func (s seq) downwave(start, sortedStart, end int) {
	if sortedStart-start == 0 {
		return
	}
	p := sortedStart + (end-sortedStart)/2
	m := s.partition(start, sortedStart, p)
	if m == sortedStart {
		if p == sortedStart {
			s.upwave(start, sortedStart-1)
			return
		}
		s.downwave(start, sortedStart, p-1)
		return
	}
	s.blockSwap(m, sortedStart, p)
	if m == start {
		if p == sortedStart {
			s.upwave(m+1, end)
			return
		}
		p++
		s.downwave(m+p-sortedStart, p, end)
		return
	}
	if p == sortedStart {
		s.upwave(start, m-1)
		s.upwave(m+1, end)
		return
	}
	s.downwave(start, m, m+p-sortedStart-1)
	s.downwave(m+p-sortedStart+1, p+1, end)
}

func (s seq) upwave(start, end int) {
	if start == end {
		return
	}
	sortedStart := end
	sortedLength := 1
	leftBound := end - 1
	length := end - start + 1
	for leftBound > start {
		s.downwave(leftBound, sortedStart, end)
		sortedStart = leftBound
		sortedLength = end - sortedStart + 1
		if length < sortedLength<<2 {
			break
		}
		leftBound = end - sortedLength<<1 + 1
	}
	s.downwave(start, sortedStart, end)
}

func (s seq) WaveSort() {
	if len(s) < 2 {
		return
	}
	s.upwave(0, len(s)-1)
}   
\end{lstlisting}

\section{wave sort - trade off}\label{appendix:trade-off}
\begin{lstlisting}[caption=Wave Sort with Trade-offs, label=lst:wavesort_tradeoff]

package trade_off

type seq []int

var compare, swap, blockSwap int

// partition splits the sequence into two parts, the left part is less than the pivot and the right part is greater than the pivot.
// It returns the index of the split point which is involved in the greater part.
// l is the index of the left bound of the unsorted part.
// r is the index of the right bound of the unsorted part which is the firt element of the sorted part.
// p is the index of the pivot.
func (s seq) partition(l, r, p int) int {
	i, j := l-1, r

	for {
		for {
			i++
			if i == j {
				return i
			}
			if s.less(p, i) {
				break
			}
		}
		for {
			j--
			if j == i {
				return i
			}
			if s.less(j, p) {
				break
			}
		}
		s.swap(i, j)
	}
}

func (s seq) less(i, j int) bool {
	compare++
	return s[i] < s[j]
}

// blockSwap swaps the block from m to r and the one from r to p.
// m is the index of the first element of the left block.
// r is the index of the right bound of the left block and the first element of the right block.
// p is the index of the last element of the right block.
func (s seq) blockSwap(m, r, p int) {
	ll := r - m
	if ll == 0 {
		return
	}

	lr := p - r + 1

	if lr == 1 {
		s[m], s[p] = s[p], s[m]
		swap++
		return
	}

	if lr <= ll {
		s.blockSwap_sr(m, r, p)
		blockSwap += lr << 1
		return
	}

	s.blockSwap_sl(m, p, ll)
	blockSwap += ll + lr
}

func (s seq) blockSwap_sl(m, p, ll int) {
	tmp := s[m]
	init := m
	j := m
	nm := p - ll + 1
	var k int
	for range p - m + 1 {
		if j >= nm {
			k = j - nm + m
			if k == init {
				init++
				s[j] = tmp
				j = init
				tmp = s[j]
				continue
			}
		} else {
			k = j + ll
		}
		s[j] = s[k]
		j = k
	}
}

func (s seq) blockSwap_sr(m, r, p int) {
	i := m
	tmp := s[i]
	j := r
	for j < p {
		s[i] = s[j]
		i++
		s[j] = s[i]
		j++
	}
	s[i] = s[j]
	s[j] = tmp
}

func (s seq) swap(i, j int) {
	swap++
	s[i], s[j] = s[j], s[i]
}

func (s seq) upwave(start, end int) {
	length := end - start + 1
	if length == 1 {
		return
	}
	if length <= 12 {
		s.insertSort(start, end)
		return
	}
	sortedStart := end - 7
	s.insertSort(sortedStart, end)
	sortedLength := 8
	leftBound := end - 127
	for leftBound > start {
		s.downwave(leftBound, sortedStart, end)
		sortedStart = leftBound
		sortedLength = end - sortedStart + 1
		if length <= sortedLength<<6 {
			break
		}
		sortedLength <<= 4
		leftBound = end - sortedLength + 1
	}
	s.downwave(start, sortedStart, end)
}

func (s seq) downwave(start, sortedStart, end int) {
	unsortedLength := sortedStart - start
	if unsortedLength == 0 {
		return
	}
	if end-sortedStart == 0 {
		if unsortedLength < 12 {
			s.insertSort(start, end)
			return
		}
		s.upwave(start, end)
		return
	}
	p := sortedStart + (end-sortedStart)/2
	m := s.partition(start, sortedStart, p)
	if m == sortedStart {
		if p == sortedStart {
			s.upwave(start, sortedStart-1)
			return
		}
		s.downwave(start, sortedStart, p-1)
		return
	}
	s.blockSwap(m, sortedStart, p)
	if m == start {
		if p == sortedStart {
			s.upwave(m+1, end)
			return
		}
		s.downwave(m+p-sortedStart+1, p+1, end)
		return
	}
	if p == sortedStart {
		s.upwave(start, m-1)
		s.upwave(m+1, end)
		return
	}
	s.downwave(start, m, m+p-sortedStart-1)
	s.downwave(m+p-sortedStart+1, p+1, end)
}

func (s seq) WaveSort() {
	if len(s) < 2 {
		return
	}
	s.upwave(0, len(s)-1)
}

func (s seq) insertSort(l, r int) {
	sortedIndex := r - 1
	for ; sortedIndex >= l; sortedIndex-- {
		low := sortedIndex
		hi := r
		m := (low + hi + 1) >> 1
		for low < hi {
			if s.less(sortedIndex, m) {
				hi = m - 1
				m = (low + hi + 1) >> 1
				continue
			}
			low = m
			m = (low + hi + 1) >> 1
		}
		if m > sortedIndex {
			tmp := s[sortedIndex]
			for j := sortedIndex; j < m; j++ {
				s[j] = s[j+1]
			}
			s[m] = tmp
			blockSwap += m - sortedIndex + 1
		}
	}
}
\end{lstlisting}

\section{wave sort - adaptive}\label{appendix:adaptiv}
\begin{lstlisting}[caption=Adaptive Wave Sort, label=lst:wavesort_adaptive]

package adaptive

type seq []int

var compare, swap, blockSwap int

// partition splits the sequence into two parts, the left part is less than the pivot and the right part is greater than the pivot.
// It returns the index of the split point which is involved in the greater part.
// l is the index of the left bound of the unsorted part.
// r is the index of the right bound of the unsorted part which is the firt element of the sorted part.
// p is the index of the pivot.
func (s seq) partition(l, r, p int) int {
	i, j := l-1, r

	for {
		for {
			i++
			if i == j {
				return i
			}
			if s.less(p, i) {
				break
			}
		}
		for {
			j--
			if j == i {
				return i
			}
			if s.less(j, p) {
				break
			}
		}
		s.swap(i, j)
	}
}

func (s seq) less(i, j int) bool {
	compare++
	return s[i] < s[j]
}

// blockSwap swaps the block from m to r and the one from r to p.
// m is the index of the first element of the left block.
// r is the index of the right bound of the left block and the first element of the right block.
// p is the index of the last element of the right block.
func (s seq) blockSwap(m, r, p int) {
	ll := r - m
	if ll == 0 {
		return
	}

	lr := p - r + 1

	if lr == 1 {
		s[m], s[p] = s[p], s[m]
		swap++
		return
	}

	if lr <= ll {
		s.blockSwap_sr(m, r, p)
		blockSwap += lr << 1
		return
	}

	s.blockSwap_sl(m, p, ll)
	blockSwap += ll + lr
}

func (s seq) blockSwap_sl(m, p, ll int) {
	tmp := s[m]
	init := m
	j := m
	nm := p - ll + 1
	var k int
	for range p - m + 1 {
		if j >= nm {
			k = j - nm + m
			if k == init {
				init++
				s[j] = tmp
				j = init
				tmp = s[j]
				continue
			}
		} else {
			k = j + ll
		}
		s[j] = s[k]
		j = k
	}
}

func (s seq) blockSwap_sr(m, r, p int) {
	i := m
	tmp := s[i]
	j := r
	for j < p {
		s[i] = s[j]
		i++
		s[j] = s[i]
		j++
	}
	s[i] = s[j]
	s[j] = tmp
}

func (s seq) swap(i, j int) {
	swap++
	s[i], s[j] = s[j], s[i]
}

func (s seq) upwave(start, end int) {
	length := end - start + 1
	if length == 1 {
		return
	}
	if length <= 12 {
		s.insertSort(start, end, end)
		return
	}
	sortedStart := s.preSorted(start, end)
	sortedLength := end - sortedStart + 1
	if sortedLength < 8 {
		s.insertSort(end-7, sortedStart, end)
		sortedStart = end - 7
		sortedLength = 8
	}
	sortedLength <<= 4
	leftBound := end - sortedLength + 1

	for leftBound > start {
		s.downwave(leftBound, sortedStart, end)
		sortedStart = leftBound
		sortedLength = end - sortedStart + 1
		if length <= sortedLength<<6 {
			break
		}
		sortedLength <<= 4
		leftBound = end - sortedLength + 1
	}
	s.downwave(start, sortedStart, end)
}

func (s seq) downwave(start, sortedStart, end int) {
	unsortedLength := sortedStart - start
	if unsortedLength == 0 {
		return
	}
	lr := end - sortedStart
	if lr == 0 {
		if unsortedLength < 12 {
			s.insertSort(start, end, end)
			return
		}
		s.upwave(start, end)
		return
	}
	p := sortedStart + lr/2
	m := s.partition(start, sortedStart, p)
	if m == sortedStart {
		if p == sortedStart {
			s.upwave(start, sortedStart-1)
			return
		}
		s.downwave(start, sortedStart, p-1)
		return
	}
	s.blockSwap(m, sortedStart, p)
	if m == start {
		if p == sortedStart {
			s.upwave(m+1, end)
			return
		}
		s.downwave(m+p-sortedStart+1, p+1, end)
		return
	}
	if p == sortedStart {
		s.upwave(start, m-1)
		s.upwave(m+1, end)
		return
	}
	s.downwave(start, m, m+p-sortedStart-1)
	s.downwave(m+p-sortedStart+1, p+1, end)
}

func (s seq) WaveSort() {
	if len(s) < 2 {
		return
	}
	s.upwave(0, len(s)-1)
}

func (s seq) preSorted(start, end int) int {
	du := false

	for i := end; i > start; i-- {
		if du {
			if s.less(i, i-1) {
				return i
			}
			continue
		}
		if s.less(i-1, i) {
			if i == end {
				du = true
				continue
			}
			s.reverse(i, end)
			return i
		}
	}
	if du {
		return start
	}
	s.reverse(start, end)
	return start
}

func (s seq) reverse(start, end int) {
	i := start
	j := end
	for range (end - start + 1) / 2 {
		s.swap(i, j)
		i++
		j--
	}
}

func (s seq) insertSort(l, sortedStart, r int) {
	sortedIndex := sortedStart - 1
	for ; sortedIndex >= l; sortedIndex-- {
		low := sortedIndex
		hi := r
		m := (low + hi + 1) >> 1
		for low < hi {
			if s.less(sortedIndex, m) {
				hi = m - 1
				m = (low + hi + 1) >> 1
				continue
			}
			low = m
			m = (low + hi + 1) >> 1
		}
		if m > sortedIndex {
			tmp := s[sortedIndex]
			for j := sortedIndex; j < m; j++ {
				s[j] = s[j+1]
			}
			s[m] = tmp
			blockSwap += m - sortedIndex + 1
		}
	}
}
\end{lstlisting}

\end{document}